\documentclass[showpacs,preprintnumbers,amsmath,amssymb,floatfix]{revtex4}
\usepackage{amsmath,graphicx,amsfonts}
\usepackage{dcolumn}

\newcommand{\be}{\begin{equation}}
\newcommand{\ee}{\end{equation}}

\begin{document}
{
\title{\bf  Conformational transformations induced by the charge-curvature interaction} 

\author{Yu. B. Gaididei}
\affiliation{Bogolyubov Institute for Theoretical Physics,
Metrologichna str. 14 B, 01413, Kiev, Ukraine}
\author{P. L. Christiansen}
\affiliation{Informatics and Mathematical Modelling and Department of Physics, 
The Technical University of Denmark, 
DK-2800 Lyngby, Denmark}

\author{W. J. Zakrzewski} 
\affiliation{Department of Mathematiccal Sciences, University of Durham, Science Laboratories, South Road, Durham DH1 3LE, England }

\date{\today}

\begin{abstract}
A simple phenomenological model for describing the conformational
dynamics of biological macromolecules via the nonlinearity-induced  instabilities is proposed. It is shown that  
 the  interaction  between charges and  bending degrees of freedom of closed molecular aggregates may act as drivers giving impetus to conformational dynamics of biopolymers.  
 It is demonstrated that  initially
circular aggregates may undergo  transformation to  polygonal
shapes and possible application to aggregates of bacteriochlorophyl $a$ 
molecules is considered. 

\end{abstract}
\maketitle


\section{Introduction}

The functioning of
biological macromolecules is determined by their
tertiary structure and  different types of semiflexible
polymers in solutions exhibit a variety of {\em conformational
phase transitions} \cite{geom}. Even modest  conformational
changes modify long-range electronic interactions in oligopeptides
 \cite{wolfgang},
 they may  remove steric hindrances and open the pathways
 for  molecular motions which  are not available in rigid  proteins
 \cite{feitel}.  In particular, it has
been recently shown \cite{viduna}  that flexibility increases the hydrogen
accessibility of DNA fragments and in this way facilitates strand breaks
in DNA molecules. There is also a strong belief that the  conductivity 
of DNA  is due to thermal motion of small polarons \cite{conw,yoo,alex}.

Research in solitonic
properties of the {\em chains with a bending} has been
initiated in the recent years
\cite{peyrard,polyakova,curv1,arch,george}.
 In particular, it was shown  that
the bending of the chain could manifest itself as an effective trap for
nonlinear excitations \cite{peyrard,curv1,george} and that the
energy of excitations decreases when the curvature of the bending
increases \cite{curv1}. 

Quite
recently,  there has been a growing interest in
 studying nonlinear  charge and energy transport  in soft condensed systems
 (polymers, membranes)
 with self-consistent account of coupling between nonlinear excitations
 and the shape of  the systems.
It has been demonstrated \cite{saxena-1998} that a mismatch of
length scales in the presence of magnetic solitons leads to an
elastic deformation on a soft  magnetic surface. A
phenomenological model for describing the conformational dynamics
of biopolymers via the nonlinearity-induced buckling and collapse
instability was proposed  in \cite{ming02}. It was shown there that the
nonlinear excitations may cause local softening of polymer bonds.
That is the effective bending  rigidity  of a chain may become
negative
 nearby the nonlinear excitation and in this way  the nonlinear excitation
causes a buckling instability of the chain.

In this paper we study a simple model for electron-curvature interactions on closed  molecular aggregates. In particular we  show that due to the   interaction between electrons and the bending degrees of freedom  the circular shape of the aggregate may be become unstable and the aggregate takes the shape of an ellipse or a polygon. 
It is shown 
that the interaction between complexes may stabilize the ringlike shape of the 
The paper is organized as follows. The theory  is applied to study the conformational transitions in light-harvesting  complexes in purple bacteria. 
In Sec. II we describe a model. In Sec. III we  
present an analytical solution to the Euler-Lagrange equation.
 In Sec.III we compare our analytical results to results 
 obtained directly by numerical simulations. In Sec. IV the interplay between  shape of complexes and  intercomplex interaction is studied. Sec. V 
 is devoted to application to  conformational transformations 
 in light-harvesting complexes.  Sec. VI presents some concluding remarks.

\section{The model}
Let us consider a  chain consisting of L units
labelled by an index $n$, and located at the points
$\vec{r}_n=\{x_n, \, y_n, \, z_n\}$. We are interested in the case
when the chain is closed and so we impose the periodicity condition on
the coordinates $\vec{r}_n$ \begin{eqnarray}\label{closed}
\vec{r}_n=\vec{r}_{n+L}.\end{eqnarray}
  The chain flexibility is accounted for by employing a
microstructure consisting of many sequentially joined rigid rods
and by incorporating a bend potential at each point of rotation
 \cite{chaubal}. Thus the Hamiltonian of
such a polymer chain has the form
\begin{eqnarray}\label{hamilt}
H=U+H_{el},\end{eqnarray}
with 
the potential energy of inter-unit interactions
$U(...,\vec{r}_n,\vec{r}_{n+1},..)$ which we take in the form
\begin{eqnarray}\label{pot}
U=U_S+U_B\end{eqnarray}
where
\begin{eqnarray}\label{potstr}
U_s=\frac{\sigma}{2}\,\sum_n\left(|\vec{r}_n-\vec{r}_{n+1}|-a\right)^2
\end{eqnarray}
is the stretching energy in the harmonic approximation. Here  $a$ is an
equilibrium distance between units (in what follows we assume $a=1$) and $\sigma$
is  a dimensionless elastic modulus of the stretching rigidity of the chain, and
\begin{eqnarray}\label{potbend}
U_b=\frac{k}{2}\,\sum_n\,\frac{\kappa^2_n}{1-\,\kappa^2_n/\kappa^2_{max}}\end{eqnarray}
is the bending energy.
Furthermore, 
\begin{eqnarray}\label{kappan}
\kappa_n\equiv |\vec{\tau}_{n}-\vec{\tau}_{n-1}|=2\,\sin\frac{\alpha_n}{2}\end{eqnarray}
determines the curvature of the chain at the point $n$.
The vector
\begin{eqnarray}\label{tang}
\vec{\tau}_n=\frac{\vec{r}_{n+1}-\vec{r}_n}{|\vec{r}_{n+1}-\vec{r}_n|}
\end{eqnarray} is the unit tangent vector,
 $\alpha_n$  is the angle between the tangent vectors $\vec{\tau}_{n}$ and $\vec{\tau}_{n-1}$, $\kappa_{max}=2\,\sin\left(\alpha_{max}/2\right)$  with  $\alpha_{max}$ being the maximum bending angle, $k$ is the elastic modulus of the bending rigidity of the chain.

We assume that there is  a small amount of extra electrons  ( or holes) on the chain.
The  Hamiltonian of electrons and their interaction with conformational degrees of freedom  is
\begin{eqnarray}
\label{ham} H_{el} = H_{hop}+H_{el-conf},\end{eqnarray}
where
\begin{eqnarray}
\label{hopp}
H_{hop}=J\,\sum_{n}\Big |  \psi_n -
\psi_{n+1}\Big |^2  \; ,
\end{eqnarray}
describes  the motion of electrons  along the chain, and  
\begin{eqnarray}
\label{el-conf}
H_{el-conf}=-\frac{1}{2}\sum_n\,\chi\,\left(|\psi_{n+1}|^2+|\psi_{n-1}|^2\right)\,
\kappa_n^2
\end{eqnarray}
gives the interaction between electrons and bending degrees of freedom. In Eqs. (\ref{ham})-(\ref{el-conf})
 $\psi_n(t)$ is the electron wave function at the site $n$, the
 parameter  $J$  describes the electron hopping in the chain, and  $\chi$ is the curvature-electron coupling constant (see Appendix for details). 
 
 The quantity
\begin{eqnarray}\label{density}\nu\equiv\frac{1}{L}\sum_n\,|\psi_n|^2\end{eqnarray}
gives the total density of extra electrons which can move along the chain and participate in the formation of the conformational state of the system. We will neglect the interaction between electrons. It is legitimate when the   total density of electrons  in the chain $\nu$ is  small.
 Combining Eqs. (\ref{potbend}) and (\ref{el-conf}),  we notice that the effective bending rigidity changes close the points where the electron is localised. For positive values of the coupling constant  $\chi$ there is a local softening of the chain, while for $\chi$ negative there is a local hardening of the chain.

 In what follows we assume that the chain is planar ($z_n=0$) and inextensible
($\sigma\rightarrow\infty$):
\begin{eqnarray}\label{inext}|\vec{r}_n-\vec{r}_{n+1}|=1.\end{eqnarray}

\section{Continuum approach}
We are interested here in the case when the characteristic size of
the excitation is much larger than the lattice spacing.  This
permits us to  replace $\psi_n(t)$  by the function $\psi(s,t)$ of
the arclength $s$ which is the continuum analogue of $n$. Using
the Euler-Mclaurin summation formula \cite{abr} we get
\begin{eqnarray}
\label{hamct}
H=U_b+H_{el},
\end{eqnarray}

\begin{eqnarray}\label{hamb}
U_b=\,\frac{1}{2}\,k\,\int\limits_{0}^{L} \,\kappa(s)^2\,d s\;
\end{eqnarray}
\begin{eqnarray}\label{hamex}
H_{el}=\int\limits_{0}^{L} \Biggl\{ J\, \Bigl|\partial_s\psi
\Bigr|^2 \,-\,\chi\,\kappa(s)^2\,|\psi|^2
\Biggr\} d s.
\end{eqnarray}
Being interested here in small curvature effects: $\kappa(s)\ll \kappa_{max}$, we consider  the bending energy in the harmonic approximation  given by Eq. (\ref{hamb}).

\subsection{Ground state of the chain}

The electron ground state wave function  $\phi(s)$  and  
the  shape of the chain $\vec{r}(s)$ may be obtained  by minimizing
the functional
\begin{eqnarray}\label{immen}
{\cal E}=H_{el}+U_b
\end{eqnarray}
with
$H_{el}$ and $U_b$ given by Eqs (\ref{hamex})
and (\ref{hamb}) under the constraint
\begin{eqnarray}\label{N}
\nu=\frac{1}{L}\int\limits_{0}^{L}\phi(s)^2\,ds,
\end{eqnarray}
which is a continuum analog of Eq. (\ref{density}). The
inextensibility constraint  (\ref{inext}) reads in the continuum limit

\begin{eqnarray}\label{inextens}
|\partial_s\vec{r}|^2=1.
\end{eqnarray}
The inextensibility constraint (\ref{inextens}) is automatically taken into
account  by choosing the parametrization
\begin{eqnarray}\label{param}\partial_s\,x(s)=\sin\theta(s),~~~\partial_s\,y(s)=\cos\theta(s)
\end{eqnarray}
where the angle $\theta(s)$ satisfies the conditions
\begin{eqnarray}\label{close1}\theta(s+L)=2\pi+\theta(s)\end{eqnarray}
and
\begin{eqnarray}\label{closure}\int\limits_0^L\,\cos\,\theta(s)\,ds=\int\limits_0^L\,\sin\,\theta(s)\,ds=0\end{eqnarray}
which follow from Eq. (\ref{closed}). Note that, in the
continuum limit, the curvature of the chain $\kappa(s)$ given by
Eq. (\ref{kappan}) can be expressed as
\begin{eqnarray}\label{kappal}
\kappa(s)=|\partial^2_s \vec{r}(s)|.
\end{eqnarray} Thus, in the frame of the parametrization (\ref{param}), $\kappa(s)=\partial_s\theta$ and the 
functional (\ref{immen}) takes the form

\begin{eqnarray}\label{energy}
{\cal E}=\int\limits_{0}^{L} \Biggl\{ J\,\nu\, \left(\partial_s\varphi
\right)^2
\,+\,\left(\frac{k}{2}-\chi\,\nu\,\varphi^2\right)\,\left(\partial_s\theta\right)^2
\Biggr\} d s,
\end{eqnarray}
where the rescaled function $\varphi(s)=\sqrt{\nu}\,\phi(s)$ which satisfies the normalization condition
\begin{eqnarray}\label{varphicond}
\frac{1}{L}\,\int\limits_0^L\,\varphi^2(s)=1,
\end{eqnarray}
has been introduced.

The Euler-Lagrange equations for the problem of minimizing ${\cal
E}$, given by Eq. (\ref{energy}) under the constraint (\ref{varphicond}) become
\begin{eqnarray}\label{eul-lag-p}
 \partial_s^2\varphi + \frac{\chi}{J}\,\left(\partial_s\theta\right)^2\,
\phi-\lambda\varphi=0,
\end{eqnarray}
\begin{eqnarray}\label{eul-lag-k}
\partial_s\left(\partial_s\theta\,\left(1-w\,\varphi^2\right)\right)=0
\end{eqnarray}
where $\lambda$ is the Lagrange multiplier and 

\begin{eqnarray}\label{w}
w=\frac{2\,\chi\,\nu}{k} 
\end{eqnarray}
is a coupling constant which characterizes the strength of the charge-curvature interaction in terms of the bending rigidity of the chain and the charge density. We are interested in solutions of Eq. (\ref{eul-lag-p}) subject to
the periodic boundary conditions
\begin{eqnarray}\label{bc}
\varphi(s)=\varphi(s+\frac{L}{n}).
\end{eqnarray}
where $n$ is an integer which characterizes the shape of the chain (see below).
Integrating Eq. (\ref{eul-lag-k}), we get
\begin{eqnarray}\label{theta}\partial_s\theta=\frac{A}{1-w\,\varphi^2}
\end{eqnarray}
where $A$ is an integration constant. 
 Taking into account the condition (\ref{close1}) we obtain that the integration constant $A$ is determined by the relation
\begin{eqnarray}\label{constA}A=\frac{2\pi}{L}\,I\end{eqnarray}
where the functional $I$ is given by the relation
\begin{eqnarray}\label{I}\frac{1}{I}=\frac{1}{L}\,\int\limits_{0}^{L}\frac{ds}{1-w\,\varphi^2}.\end{eqnarray}

 From Eqs. (\ref{param}) and (\ref{theta}) we see that the shape of the chain is determined by the equations
\begin{eqnarray}\label{shape}
x(s)=\int\limits_0^s\,sin\,\theta(s')\,ds',~~~
y(s)=\int\limits_0^s\,cos\,\theta(s')\,ds',\nonumber\\
\theta(s)=\frac{2\pi}{L\,I}\int\limits_0^s\,\frac{1}{1-w\,\varphi^2(s')}\,ds'.
\end{eqnarray}

\subsection{Solution of the Euler-Lagrange equations}

 There are two kinds of solutions to
Eqs (\ref{eul-lag-p}) and (\ref{theta}).
\begin{itemize}
\item  Circular chain.\\Charge is  uniformly distributed along the chain
\begin{eqnarray}\label{phic}\varphi=1\end{eqnarray}
where the normalization condition (\ref{N}) has been used, and the
curvature of the chain is constant
\begin{eqnarray}\label{curv}\kappa(s)\equiv\partial_s\theta=\frac{A}{1-w}.\end{eqnarray}
This case corresponds to a circular chain
$$x=R\,\sin\frac{s}{R},~~y=-R\cos\frac{s}{R}.$$ The radius $R$ of
the circle can be obtained by puting Eq. (\ref{curv}) into
the boundary condition (\ref{close1}). As a result we have
\begin{eqnarray}\label{radius}R=\frac{L}{2\pi}.\end{eqnarray} 
The energy of the circular chain is thus 
\begin{eqnarray}\label{encirc}{\cal E}_{circ}=\frac{2\pi^2}{L}\,k(1-w).\end{eqnarray}

\item Polygonally deformed chain.\\
Let  us consider now the case of  spatially non-uniform
distributed electrons. Inserting Eqs. (\ref{theta})-(\ref{I}) into Eq. (\ref{energy}) we get
\begin{eqnarray}\label{energymod}
{\cal E}=J\,\nu \,\int\limits_{0}^{L}  \left(\partial_s\varphi
\right)^2\,ds
\,+\frac{2\pi^2 k}{L}\,I.
\end{eqnarray}
We restrict our analytical consideration to the case when the  charge-curvature coupling is weak and/or the  charge density is low: $w\ll\, 1$.
Expanding the functional $I$ in terms of the small parameter $w$ we obtain from Eq. (\ref{energymod})
\begin{eqnarray}\label{energyappr}
{\cal E}=\frac{2\pi^2 \,k}{L\,(1+w)}+J\,\nu\,\int\limits_{0}^{L}  \Big\{\left(\partial_s\varphi
\right)^2
-\frac{G\nu}{R^2}\,\varphi^4-w\,\frac{G\nu}{R^2}\,\varphi^6\Big\}\,ds\,
\end{eqnarray}
where 
\begin{eqnarray}\label{g}
G=\,\frac{2\chi^2}{J\,k\,(1+w)^2}
\end{eqnarray}
is an effective nonlinear parameter.  For small $w$ one can neglect the last term in Eq. (\ref{energyappr}) and the Euler-Lagrange equation for the functional (\ref{energyappr}) then takes the form
\begin{eqnarray}\label{eqvarphi}
\partial_s^2\varphi+2 \frac{G\,\nu}{R^2}\,\varphi^3-\lambda\,\varphi=0.
\end{eqnarray}
 Straightforward  calculations 
show that Eq. (\ref{eqvarphi})  has a  solution of the form 
\begin{eqnarray}\label{varphisol}
\varphi=R\,\sqrt{\frac{\lambda}{(2-m)\,G\,\nu}}\,dn\left(\sqrt{\frac{\lambda}{(2-m)}}\,s\Big | m\right)
\end{eqnarray}
 where
$\,dn(u |m)$ is the Jacobi  elliptic function with the modulus $m$ \cite{abr}. Inserting Eq. (\ref{varphisol}) into the boundary condition (\ref{bc}) and the normalization condition (\ref{varphicond}), we find that the Lagrange multiplier $\lambda$ and the modulus $m$  are determined by the equations
\begin{eqnarray}\label{lambda}\sqrt{\frac{\lambda}{(2-m)}}\,\frac{L}{n}=2\,K(m)
\end{eqnarray}

\begin{eqnarray}\label{meq}G\,\nu=\frac{n^2}{\pi^2}\,K(m)E(m)\end{eqnarray}
and the charge distribution along the chain is given by
\begin{eqnarray}\label{chargedistr}
\varphi=\,\sqrt{\frac{K}{E}}\,dn\left(\frac{2\,n\,K}{L}\,s\Big | m\right)
\end{eqnarray}
where $K(m)$ and $E(m)$ are  the complete elliptic integrals of the first kind and the second kind,  respectively \cite{abr}.

The curvature of the chain is given by the equation
\begin{eqnarray}\label{curvature}\kappa(s)\equiv\partial_s\theta
\approx\,\frac{1}{R}\,I\,\left(
1+w\,\frac{K}{E}\,dn^2\left(\frac{2n\,s\,K}{L}\Big|m\right)\right).
\end{eqnarray} 
 
Integrating Eq.
(\ref{curvature}), we get
\begin{eqnarray}\label{thetaE}\theta(s)=\,\frac{2 \pi}{L}\,I\,\left(
s+w\,\frac{L}{2\,n\,E}\,E\left(\alpha\big|m\right)\right)\end{eqnarray} 
where $E\left(\alpha\big|m\right)$ is the incomplete elliptic integral of the second kind, and $\alpha=am\left(\frac{2\,n\,K}{L}\,s\big |m\right)$ is the amplitude function \cite{abr}. By using the Fourier expansion for the amplitude function for small $m$ we obtain from Eq. (\ref{thetaE})

\begin{eqnarray}\label{thetas}\theta(s)\approx\frac{2\pi}{L}\,(1+w)\,I\,s+
\frac{w}{4 n }\,I\,m\,\sin\left(\frac{2 n \pi}{L}\,s\right).\end{eqnarray}

Inserting Eq. (\ref{thetas}) into the closure condition  (\ref{closure}), we find that it is satisfied  for  $n\geq 2$. 
Eqs. (\ref{shape}) and (\ref{thetas}) describe  a polygon: for $n=2$  it is an elliptically deformed chain, while for $n=3$ it has a triangular shape (see Fig.~\ref{figshape} ). We see from Eqs.  (\ref{chargedistr}) and (\ref{curvature}) that the polygon structure is a result of the self-consistent interaction between electrons and bending degrees of freedom: extrema of the curvature and of the charge density correlate: in the case of the softening electron-curvature interaction ($\chi>0$ maxima  of curvature and charge density coincide, while in the case of the hardening interaction the minima of the curvature coincide with the maxima of the charge density.   Eq. (\ref{meq}) shows that, for a given value of the nonlinear parameter $G$, the $n$-gon structure  appears when the charge density exceeds the threshold value $\nu_n$: \begin{eqnarray}\label{nun}\nu\,>\nu_n\equiv\,\frac{n^2}{4\,G}\end{eqnarray}. 

The energy difference between the $n$-gon structure and  the circular  chain is given by the expression
\begin{eqnarray}\label{energydiff}
{\cal E}_n-{\cal E}_{circ}=\frac{4\pi^2}{ 3\,L}\,\frac{G\,J\,\nu^2}{E^2}\,\left(3 E^2-(2-m) E K-(1-m) K^2\right).
\end{eqnarray}
The normalized energy difference
\begin{eqnarray}\label{delta}
\Delta_n=\frac{{\cal E}_n-{\cal E}_{circ}}{{\cal E}_{circ}}
\end{eqnarray}
for $n=2,3$  versus the charge density is shown in Fig. \ref{figenergy}. We note that  when the charge density is above the critical value   the deformed structure  with spatially inhomogeneous  charge distribution is energetically more favorable than the circular system with a uniformly distributed charge. The state with elliptically deformed chain $n=2$ is the  energetically most preferable.

Note also that our analytical approach  was based on the assumption that $w\,\varphi^2\,\ll\,1$. Taking into account Eq. (\ref{chargedistr}), this means that it is legitimate to consider not too sharp distributions which correspond to $w\,K(m)\ll 1$ or $m\,\ll  1-\exp\{-0.72/w\}.$

\end{itemize}

\section{Numerical studies}
To check our results we have performed also several numerical studies. To this end we carried out the dynamical simulations of the  equations
\begin{eqnarray}\label{dyneqr}
\eta\frac{d}{dt}\vec{r}_n=-\frac{\partial H}{\partial\,\vec{r}_n},
\end{eqnarray}

\begin{eqnarray}\label{dyneqpsi}i\,\frac{d}{dt}\psi_n= -\frac{\partial\, H}{\partial\,\psi^*_n}
\end{eqnarray}
 with  the Hamiltonian $H$ being defined by Eqs. (\ref{hamilt})-(\ref{el-conf}). Thus the conformational dynamics is considered in an overdamped regime with  the friction coefficient $\eta$. Then we took as our starting
configurations systems involving the electric charge density of (almost) the same magnitude ($\psi_n$)
at all points (we broke the symmetry by increasing the density at one point of the chain by $1\%$).
Initially, all the lattice points were placed at symmmetric points on the circle of an appropriate
radius (see Fig.\ref{figshapechargecurvinitial}). 
We performed such simulations for several values of the charge density. Due to the absorption 
the energy of the system was decreasing during the evolution and the system was evolving towards a minimum.
At the same the points of the chain were moving from their initial to their new positions.

We considered   both the cases of the hardening and of the softening electron-curvature interaction.

 In the case of a hardening electron-curvature interaction  ($\chi\,<\,0$)  a typical final distribution of the chain points is shown 
in Fig.\ref{figshapechargecurvhard}.  The charge density  and the curvature distributions are in full agreement with the results of our analytical considerations:  the curve is more flat where the density of the electrons is maximal. In fact, the ellipse-like  shape is rather robust as it arises
for a large range of parameters (of the strength of the hardening electron-curvature interaction 
and of the anharmonicity coefficient $\kappa_{max}$).  

 Complexes with a softening electron-curvature interaction are much more flexible. Their equilibrium shape depends drastically both on the anharmonicity and on the charge density. Figs \ref{figshapechargecurvsoftinterm} and \ref{figshapechargecurvsoftfinal} demonstrate how drastically  the shape of the complex and  the charge distribution along the chain  can change as a function of the total charge density $\nu$:    increasing the total charge  by $5\%$  can lead to the localisation of almost the whole charge of the system at one place.

In our numerical work we also studied the stability of our `final' field configurations - i.e. the configurations
which we thought the system was settling at. 
This we studied by perturbing the system. Such perturbations were
introduced in two stages. First we changed the electric charge 
of the configuration by multiplying all `final' values of the electric charge 
by a constant factor $\mu$; this had the effect of changing the energy of the system. Then 
we performed the new minimisation and, when the system appeared to have 
settled at the new `final' configuration,  we changed  back its $\psi_n$ 
 by a new multiplication by ${1/ \mu}$. As $\sum_n \vert \psi_n\vert^2$ is conserved
during the evolution, the final system had the same value of it
as the original 'unperturbed' fields. The results
of the further minimisation were then compared with the original `final' fields.

When we applied this technique to our field configuration shown in Fig. \ref{figshapechargecurvsoftfinal}
 we found that the system
was really unchanged by this perturbation; in fact the perturbation led to an overall rotation
of the system by one lattice point, but the sequence of values
of the fields was essentially the same thus showing the stability of the found minimum.

\section{Effects of intercomplex interaction}
The aim of this section is to investigate how the interaction between complexes  influences   the shape. We will consider 
the system which is described by the Hamiltonian
\begin{eqnarray}\label{hamcomp} {\cal H}=\sum_j {\cal E}_j+\frac{1}{2}\,\sum_{i,j} U_{i\,j},\end{eqnarray}
where 
${\cal E}_j$ is the energy of the $j$-th aggregate which is given by Eq. (\ref{energy}) with $\varphi$ replaced by $\varphi_j$ and $\theta$ replaced by $\theta_j$, and $U_{ij}$ is the interaction energy between particles. The latter we will take in the form of the Gay-Berne potential \cite{gayberne} which is a generalization of the Lennard-Jones 12-6 potential and  is widely used to study translational and orientational ordering in systems of aspherical molecules.  We consider small deviations from the ringlike structure of aggregates and so we neglect the difference of  the well depths for side-to-side  and end-to-end configurations. In this case the Gay-Berne potential between two parallel uniaxial molecules is given by

\begin{eqnarray}\label{gb}U_{ij}\equiv U\left(\vec{r}_{ij}\right)=\frac{U_0}{\sqrt{1-\zeta^2}}\,\Big[\left(\frac{\sigma_0}{r_{ij}+\sigma\left(\hat{\vec{r}}_{ij}\right)+\sigma_0}\right)^{12}-\left(\frac{\sigma_0}{r_{ij}+\sigma\left(\hat{\vec{r}}_{ij}\right)+\sigma_0}\right)^{6}\Big]\end{eqnarray}
where 
$\vec{r}_{ij}=r_{ij}\hat{\vec{r}}_{ij}$ is the interparticle vector, \begin{eqnarray}\label{anisotropy}\sigma\left(
\hat{\vec{r}}_{ij}\right)=\sigma_0\,\Big[1-\frac{2\zeta}{1+\zeta}\,
\left(\hat{\vec{r}}_{ij}\cdot\vec{e}\right)^2\Big]^{-1/2}\end{eqnarray}
is the anisotropy parameter where $\vec{e}~$ is a unit vector specifying the axes of symmetry. The anisotropy coefficient $\zeta$ is determined by the lengths of the major   and minor axes $\sigma_{||}$ and $\sigma_{\perp}$
\begin{eqnarray}\label{zeta}\zeta=\frac{\sigma^2_{||}-\sigma^2_{\perp}}{\sigma^2_{||}+\sigma^2_{\perp}}\end{eqnarray}
and $\sigma_0$ gives a charasteristic length scale while  $U_0$ determines the intensity of the interaction. 

The centers of densely packed circular aggregates of the radius $R$  create a two-dimensional triangular lattice $\vec{r}_{j}=j_1 \vec{a}_1+j_2 \vec{a}_2,~~~(j\equiv (j_1,j_2), ~j_1,j_2=0,\pm 1,\pm 2,...)$   with the basic vectors $\vec{a}_1=\ell\,\left(1,0\right)$ and $\vec{a}_2=\ell\,\left(1/2,\sqrt{3}/2\right)$ where $\ell$ is the lattice constant (see Fig. \ref{figcirclepack}). Being elliptically deformed in such a way that the major axes of all aggregates are parallel to the $x$-axis, the centres of densely packed elliptical aggregates create a lattice $\vec{r}_{j}=j_1 \vec{b}_1+j_2 \vec{b}_2$  with the basic vectors $\vec{b}_1=\ell\,\left(1+u,0\right)$ and $\vec{b}_2=\ell\,\left(1/2 \,(1+u),\sqrt{3}/2\, (1-u)\right)~$ (see Fig. \ref{figellipsepack}),   where  the parameter $u$ is given by
\begin{eqnarray}\label{mma}u=\frac{\sigma_{||}- \sigma_{\perp}}{\sigma_{||}+ \sigma_{\perp}}.\end{eqnarray}

Now we study the ground state of this system  by using a trial function approach. We assume that this
state 
 is spatially homogeneous and relying on the results of the previous section, we
assume that the electron trial function and the trial curvature can be taken in the form 
\begin{eqnarray}\label{trialfunction}\varphi_j=\,\left(\cos\alpha+\sqrt{2}\,\sin\alpha\,\cos\left(\frac{2\, s}{R}\right)\right),\end{eqnarray}

\begin{eqnarray}\label{trialcurvature}\partial_s\,\theta_j=\frac{1}{R}
\left(1+\gamma \,\cos\left(\frac{2\,s}{R}\right) \right)\end{eqnarray}
 The functions (\ref{trialfunction}) and (\ref{trialcurvature}) can be considered as a truncated Fourier expansion of the  solutions (\ref{varphisol}) and (\ref{curvature}) in which the coefficients $\alpha$ and $\gamma$ are  variational parameters and $R$ is the radius of the cylindrically symmetric aggregate. The function (\ref{trialfunction}) satisfies both the periodicity condition (\ref{bc}) and  the number of particles constraint (\ref{N}).  In the limit   $~\gamma < 1~$ the shape of the curve with the curvature given by Eq. (\ref{trialcurvature}) is parametrically determined by the expressions
\begin{eqnarray}\label{trialshape}x(s)=R\,\left((1+\frac{\gamma}{4})\,\cos\left(\frac{s}{R}\right)+\frac{\gamma}{12}\,\cos\left(\frac{3\,s}{R}\right)\right),\nonumber\\
y(s)=R\,\left((1-\frac{\gamma}{4})\,\sin\left(\frac{s}{R}\right)+\frac{\gamma}{12}\,\sin\left(\frac{3\,s}{R}\right)\right).
\end{eqnarray}
Thus the lengths of the major and minor axes of the curve (\ref{trialshape}) are given by 
\begin{eqnarray}\label{trialmma}\sigma_{||}=R\,(1+\frac{\gamma}{3}),~~~~\sigma_{\perp}=R\,(1-\frac{\gamma}{3})\end{eqnarray}
 and comparing Eqs (\ref{mma}) and (\ref{trialmma}), we see that $u=\gamma/3.$

Inserting Eqs. (\ref{trialfunction})  and   (\ref{trialcurvature}) into Eqs. (\ref{hamcomp}),  (\ref{energy}) and (\ref{gb}) for an energy per aggregate we get
\begin{eqnarray}\label{entrial} \frac{{\cal H}}{N_a}={\cal E}_{tr}+U\end{eqnarray}
where
\begin{eqnarray}\label{ensing}{\cal E}_{tr}=\frac{\pi\,k}{R}\Bigg\{1+\frac{\gamma^2}{2}-\frac{1}{8}\,\xi\nu\,\left(8+5\gamma^2-\gamma^2\,\cos(2\,\alpha)+8\sqrt{2}\,\sin(2\,\alpha)\right)+\frac{4\,J}{k}\,\nu\,\sin^2\alpha\Bigg\}\end{eqnarray}
is the energy of an isolated aggregate, and 
\begin{eqnarray}\label{int}U=\sum_{j=0}^5\,U\left(\vec{\Delta}_j\right)\end{eqnarray}
is the  energy due to the interaction between aggregates in the lattice.
In Eq. (\ref{int})  the function $U\left(\vec{\Delta}_j\right)$ is given by Eqs. (\ref{gb}), (\ref{anisotropy}) with  $\vec{e}=\left(1, ~0\right)$ and  vectors $\vec{\Delta}_j=\sigma_0\,\left((1+\frac{\gamma}{3})\,\cos\left(\frac{\pi j}{3}\right),(1-\frac{\gamma}{3})\,\sin\left(\frac{\pi j}{3}\right)\right)$ connect nearest and next-nearest neighbours of the lattice. The interaction energy (\ref{int}) has a minimum at $\ell=2^{1/6}\,\sigma_0,~~~\gamma=0$ which corresponds to a system of densely packed circular aggregates. Expanding the function (\ref{int}) in the vicinity of this point, in powers of the variational parameter $\gamma$,  we get
\begin{eqnarray}\label{Uexp} U=-\frac{3}{2}\,U_0+c\,U_0\,\gamma^2+\cdots\end{eqnarray}
where the numerical coefficient $c\approx 2.05.$ According to the variational principle we should satisfy the equations
\begin{eqnarray}\label{vareqa}\partial_{\alpha}\,{\cal E}_{tr}=0,\end{eqnarray}\begin{eqnarray}\label{vareqg}\partial_{\gamma}\,\left({\cal E}_{tr}+U\right)=0.\end{eqnarray}
From Eq. (\ref{vareqa}) we  get
\begin{eqnarray}\label{varalpha}\tan(2\alpha)=\frac{8\sqrt{2}k\,\xi}{32 \,J-k\,\xi\,\gamma^2}\,\gamma.\end{eqnarray}
Inserting Eq. (\ref{varalpha}) into Eq. (\ref{ensing}) and expanding it in terms of   $\gamma$ we obtain that
\begin{eqnarray}\label{ensingser}{\cal E}_{tr}={\cal E}_{circ}+\frac{\pi\,k}{2\,R}\left(1-\frac{\nu}{\nu_{cr}}\right)\,\gamma^2+\,B\,\gamma^4+\cdots
\end{eqnarray}
where 
\begin{eqnarray}\label{nucr}\nu_{cr}=\frac{k}{2\,\chi}\,\frac{J}{J+\chi}\end{eqnarray}
is the critical charge density
and the notation $$B=\frac{\pi\,}{16\,R}\frac{\chi^3\,\nu}{\,J^3}\,(2\chi-J)$$
 is introduced.   Thus in the framework of the variational approach the energy of a single complex has a single minimum at $\gamma=0$ when $\nu\,<\,\nu_{cr}$ and in this case the aggregate has a ring-like shape. When  $\nu\,>\,\nu_{cr}$  and $2\,\chi>J$ the energy (\ref{ensingser}) possesses two equivalent minima with $\pm \gamma_0\, ~(\gamma_0\neq 0$) ( see  Fig. \ref{figeffenergy}). As it is seen from Eqs. (\ref{trialshape}) the finite value of $\gamma_0$ implies that the aggregate is elliptically deformed either along the $x$-axis (when $\gamma > 0$) or along the $y$-axis (when  $\gamma < 0$). 
    Note that in the limit $\chi\gg J$,  $\nu_{cr}$ coincides with $\nu_2$ given by Eq. (\ref{nun}). Combining (\ref{Uexp}) and (\ref{ensingser}) we see that  the interaction between aggregates modifies the condition for appearance of the low-symmetry form. Indeed, even in the case when $\nu>\nu_{cr}$ (an isolated aggregate has an ellipse-like shape) in the condensed phase of aggregates for strong enough inter-aggregate interaction we may have an inequality
\begin{eqnarray}\label{ineqsymm}\nu <\,\nu_{cr}\,\left(1+\frac{2\,c \,U_0}{\pi\,k}\,R\right)\end{eqnarray}
which means that interacting aggregates are ring-like and create a densely packed crystallic structure with the group symmetry $D_{6h}$ (see Fig. \ref{figcirclepack}).When
\begin{eqnarray}\label{ineqasymm}\nu\, >\,\nu_{cr}\,\left(1+\frac{2\,c \,U_0}{\pi\,k}\,R\right)\end{eqnarray}
the ellipse-like shape survives in the condensed phase which has a rectangular unit cell and transforms in accordance with the group symmetry $D_{2h}$ ( see Fig.\ref{figellipsepack}). There are two equivalent types of arrangement in the low-symmetry phase when the aggregates are elliptically deformed either along the $x$-axis (when $\gamma > 0$) or along the $y$-axis (when  $\gamma < 0$). In accordance with this the condensed phase of the aggregates must have a domain structure, {\it i.e.} it must consist of various regions in which the direction of long axes  are different. 

\section{Application to bacteriochlorophyll $a$ complexes}
 The X-ray crystallography shows \cite{dermott}  that  bacteriochlorophyll a (Bchl a) molecules in the LH complex are organized in two concentric rings: the B800 and B850 rings. The former  consists of nine well-separated Bchl a molecules with an absorption band at ~800 nm and the latter consists of 18 Bchl  molecules with an absorption band at ~860 nm (see Fig. 1 in Ref. \cite{kete}). Measurements of the anisotropic properties of the absorption  of isolated LH2 complexes from  bound to mica surfaces \cite{bopp}, the fluorescence-excitation spectra of individual
LH2 complexes from bacteria {\it Rhodopseudomonas acidophila}  \cite{oijen,kete,matsu} showed that the complexes are generally not cylindrically symmetric but reveal a deformation of the circular complex into a shape with $C_2$ symmetry.  Quite recently the shape of the LH2 complex from bacteria
{\it Rhodobacter spheroides 2.4.1} in detergent solution has been determined from synchrotron small-angle X-ray scattering data \cite{hong}. It was shown that, in contrast to the cylindrical crystal structure with a diameter of $6.8 nm$, the shape of an isolated LH2 complex is an oblate plate with an eccentricity $\epsilon=0.59$. It was conjectured in Ref. \cite{oijen} that the extremely dense packing of LH2 in the crystals causes the  cylindrical symmetry of the complexes. A variety  in shapes and conformations for light-harvesting LH1 complexes and different types of their packing  in two-dimensional crystals revealed by atomic force microscopy were reported recently in Ref. \cite{bahatyrova}.

A qualitative explanation of this phenomena follows from the theory developed in previous sections.
It has shown that  in the presence of sufficiently strong   charge-curvature interaction an isolated complex has an ellipse-like shape while  the interaction between complexes in the form of an anisotropic Gay-Berne potential  stabilizes the ringlike shapes of the complexes. However if the intensity of the intercomplex interaction is less than some threshold value (see Eqs. (\ref{Uexp}) and (\ref{ensingser})) the non-circular shape of the complex is preserved in the condensed phase.

The  model is too simplified for quantitative predictions for light-harvesting complexes but we believe that it contains interesting physics which should be important for further studies of such systems.

\section{Conclusions}
In this paper, we have investigated the role of the electron-curvature interaction on the formation of the ground state of  closed semiflexible  molecular chains. We have found that the coupling between electrons and the bending degrees of freedom of the chain can induce a local softening or hardening of  chain bonds, {\it i.e.} the effective bending rigidity of the semiflexible chain changes as the density of charge changes along the chain. When the  charge density  and/or the strength of the electron-curvature coupling exceed a  threshold value,  the spatially uniform distribution of the charge along the chain and the circular, cylindrically symmetric shape  of the chain  become unstable. In this case the ground state of the system is characterized by a spatially non-uniform distribution of electrons along the chain and the chain takes on  an ellipse-like (or in general polygon-like) form.

\section*{Acknowledgments}
Yu. B. G. would like to thank L. Valkunas for attracting his attention to the problem of structural deformations of light-harvesting complexes. Yu.B.G. and W.J.Z. thank L. Brizhik, A. Eremko and B. Piette for their interest and 
helpful  discussions.  Yu.B.G. thanks for a Guest Professorship  funded by Civilingeni{\o}r  Frederik Christiansens Almennyttige Fond  and MIDIT, SNF grant $\#$  21-02-0500 as well as support from the research center of quantum medicine "Vidguk". He also thanks Department of Physics, the Technical University of Denmark for hospitality. 

\appendix                      
\section{} \label{appendA} 
The electron-conformation interaction  can be obtained  from the Coulomb interaction between an electron and  charged groups of the molecule   

\begin{eqnarray}\label{el-conf-app}
H_{int}=\sum_{n,n'}V\left(|\vec{r}_n-\vec{r}_{n'}|\right)\,|\psi_n|^2\end{eqnarray}
where the matrix element $V\left(|\vec{r}_n-\vec{r}_{n'}|\right)$ describes the Coulomb interaction between an electron which occupies the site $n$ in the chain and  the charged group  at the site $n'$. From  the inextensibility condition (\ref{inext}) we see that  the nearest neighbour interaction terms  ($|n-n'|=1$) do not contribute to the electron-conformational interaction. Taking into account that
\begin{eqnarray}\label{rdif} \left(\vec{r}_{n+2}-\vec{r}_n\right)^2=2 a^2+2\,\left(\vec{r}_{n+1}-\vec{r}_n\right)\cdot \left(\vec{r}_{n+2}-\vec{r}_{n+1}\right)=4 a^2-a^2\kappa_{n+1}^2\end{eqnarray}
we obtain from Eq. (\ref{el-conf-app})  that, in the next-nearest-neighbour approximation,  the energy of the electron-conformational interaction takes the form
\begin{eqnarray}\label{electr-conf}
H_{el-conf-app}=-\frac{1}{2}\sum_n\,\chi\,\left(\kappa_{n+1}^2+\kappa_{n-1}^2\right)\,|\psi_n|^2
\end{eqnarray}
where  the notation
\begin{eqnarray}\label{chi}\chi=\frac{1}{2}\,a\,\frac{dV}{dr}\Big |_{r=2a}\end{eqnarray}
has been used.

\newpage
\begin{figure}
\includegraphics[scale=1.1]{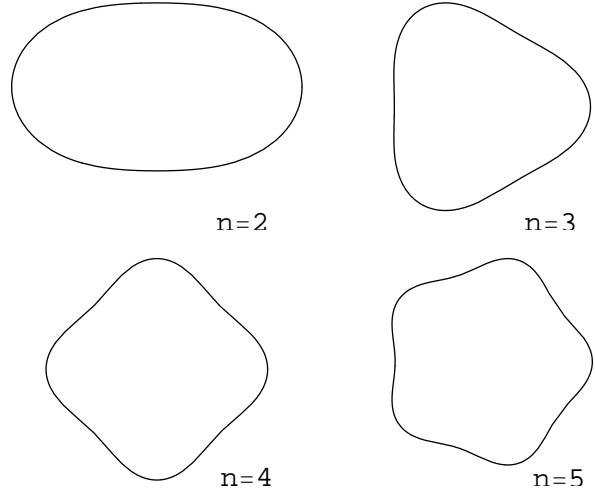}
\caption{\label{figshape} The shape of the chain:  in the ellipse-like state ($n=2$),  in the polygon states ($n=3,4,5$).}
\end{figure} 

\begin{figure}
\includegraphics[scale=1.2]{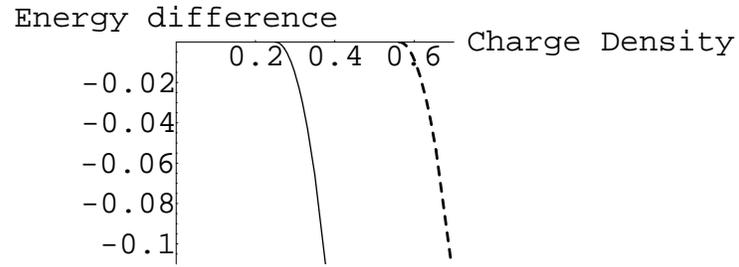}
\caption{\label{figenergy} The normalized energy difference  $\Delta_n$  from Eq. (\ref{delta} for  the  ellipse-like $n=2$  ( solid line ) and  triagon-like chain $n=3$ (dashed line)  with   $~2\chi=4\,J=k$.}
\end{figure}

\newpage
\begin{figure}
\includegraphics[scale=1]{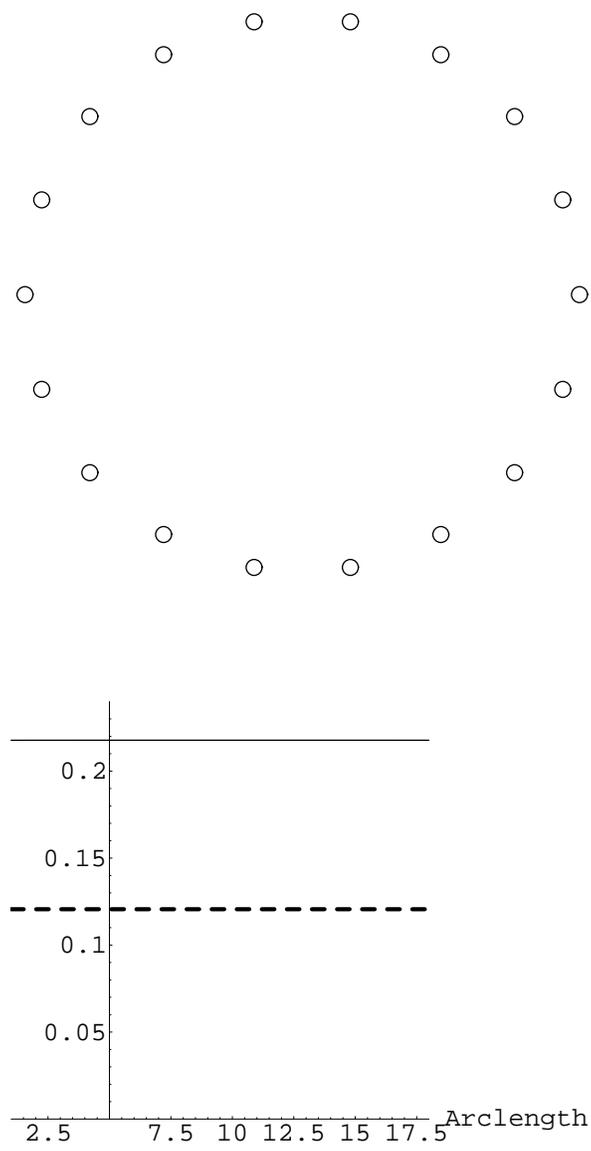}
\caption{\label{figshapechargecurvinitial} The top panel shows the initial shape of the chain; the bottom panel shows the initial charge  (solid line) and curvature (dashed line) distribution along the chain.}
\end{figure}
\newpage
\begin{figure}
\includegraphics[scale=1]{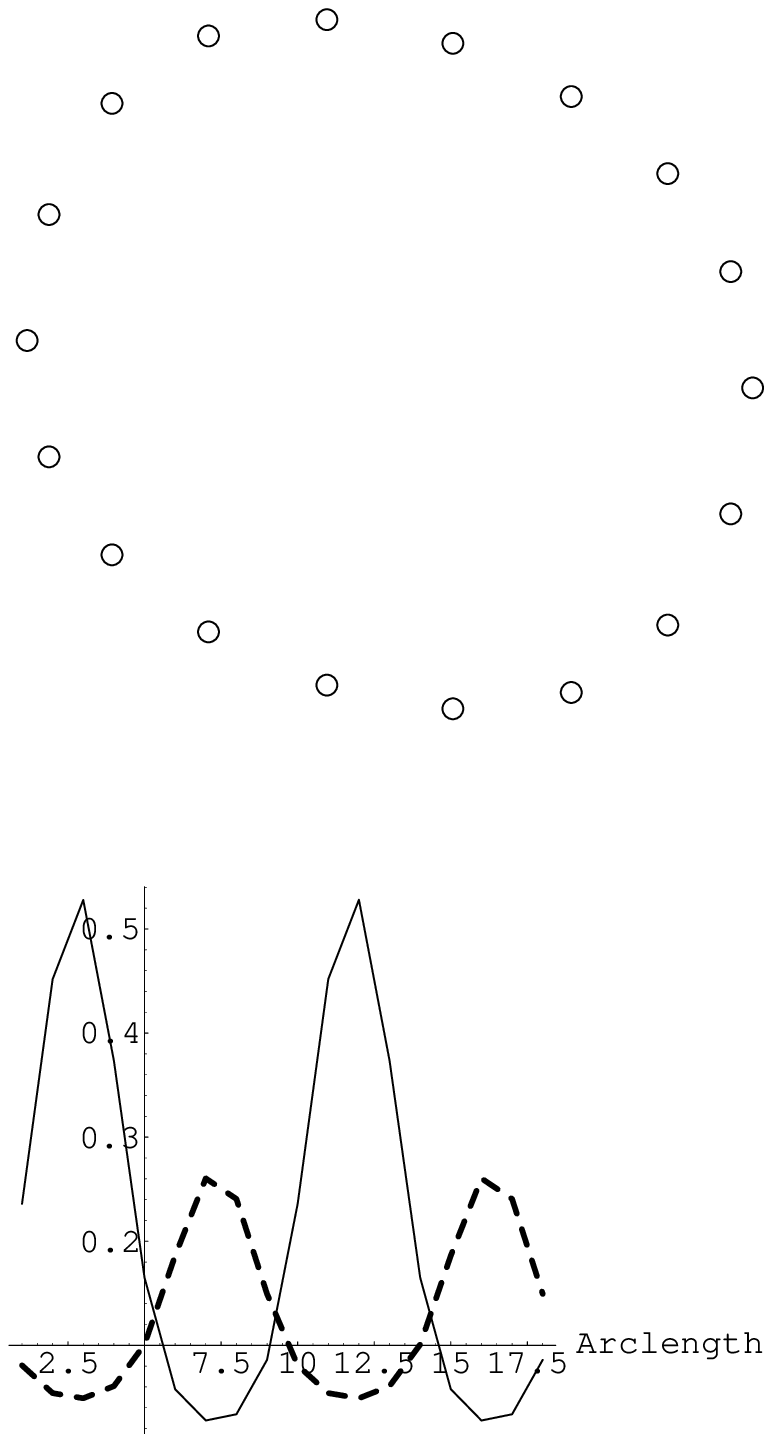}
\caption{\label{figshapechargecurvhard} The top panel shows the equilibrium shape of the chain, and the bottom panel shows the  charge distribution (solid line) and curvature variation (dashed line) along the chain  in the case of hardening electron-curvature interaction with $\nu=0.22,~\chi=-2,\,k=1 \, J=0.4\,$}
\end{figure}
\newpage
\begin{figure}
\includegraphics[scale=1]{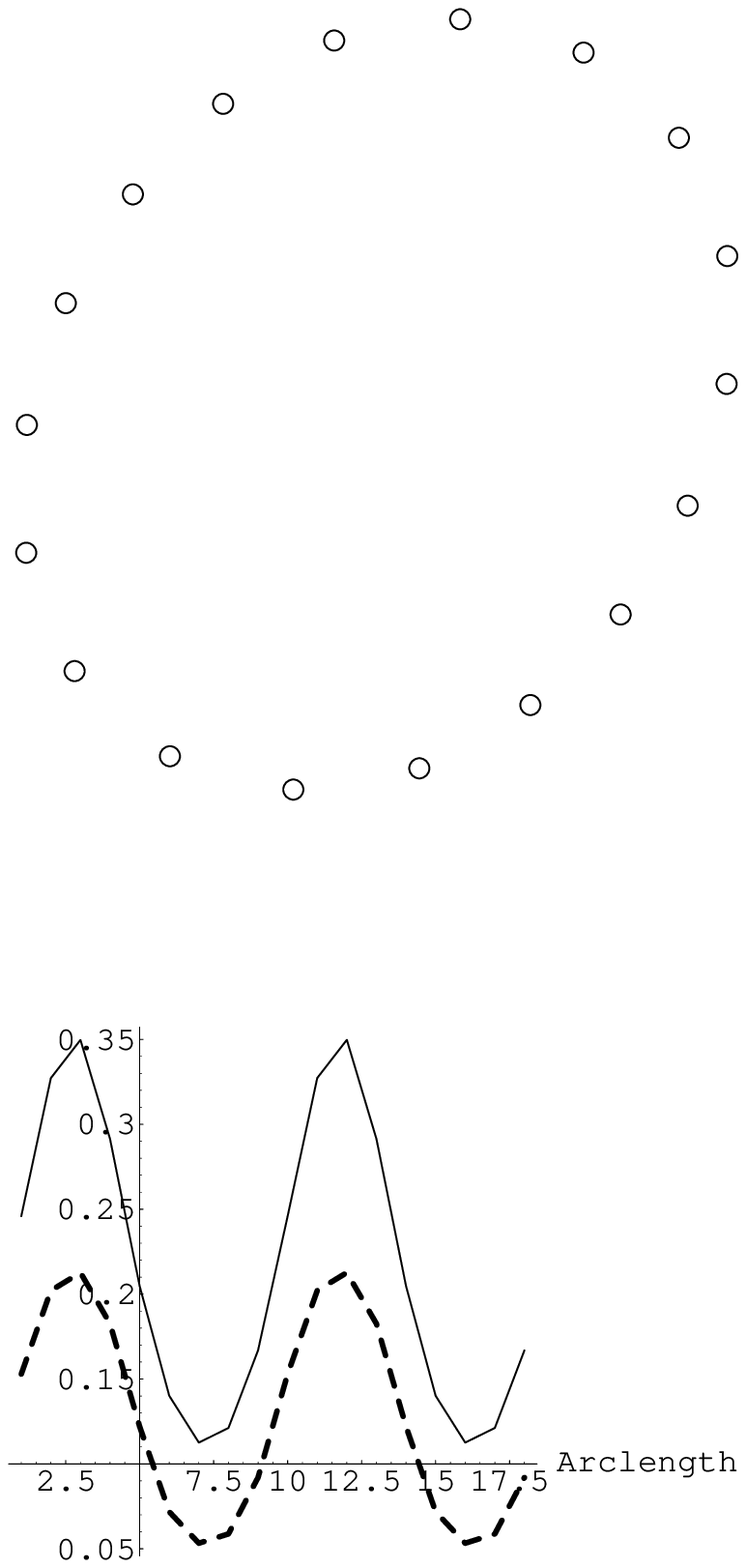}
\caption{\label{figshapechargecurvsoftinterm} The top panel shows the equilibrium shape of the chain and the bottom panel shows the  charge distribution (solid line) and curvature variation (dashed line) along the chain  in the case of softening electron-curvature interaction with $\nu=0.22,~\chi=
2.15 , \,k= J=1$}
\end{figure}
\newpage
\begin{figure}
\includegraphics[scale=1]{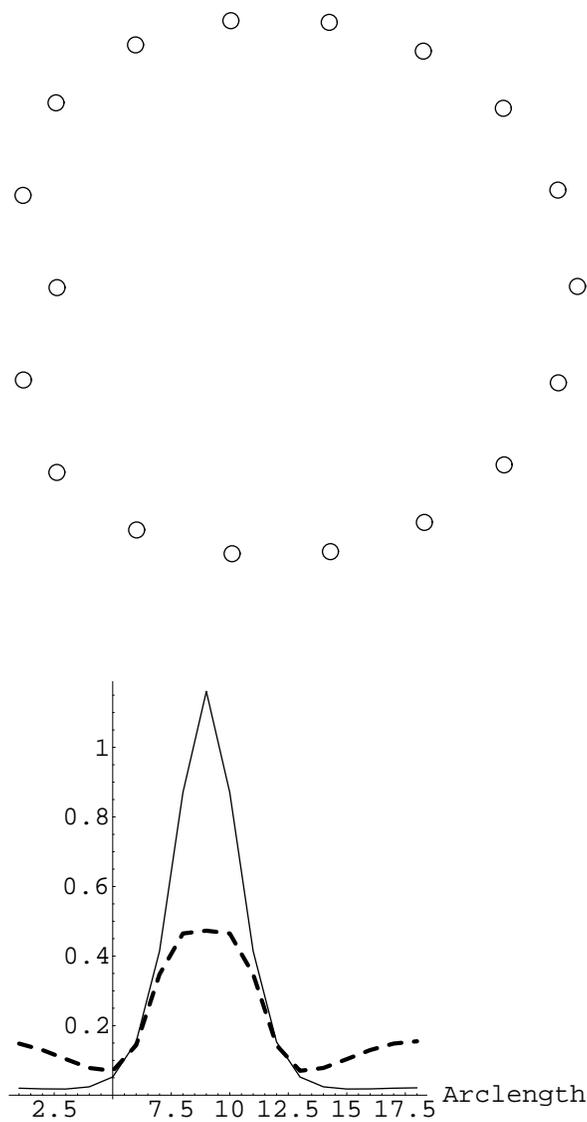}
\caption{\label{figshapechargecurvsoftfinal} Same as Fig. \ref{figshapechargecurvsoftinterm} with $\nu=0.23,~\chi=2.15 , \,k= J=1$}
\end{figure}

\begin{figure}
\includegraphics[scale=1]{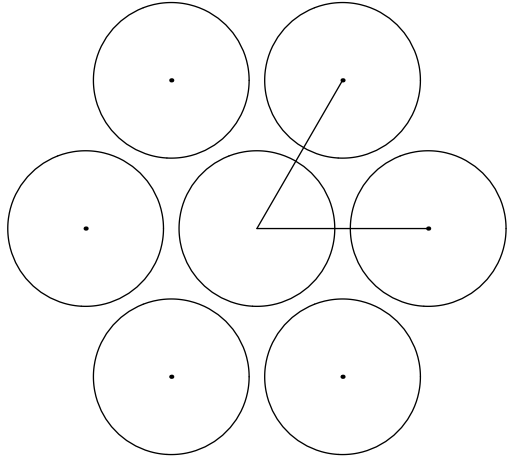}
\caption{\label{figcirclepack} Arrangement of densely packed ring-like aggregates.}
\end{figure}
 \begin{figure}
\includegraphics[scale=1]{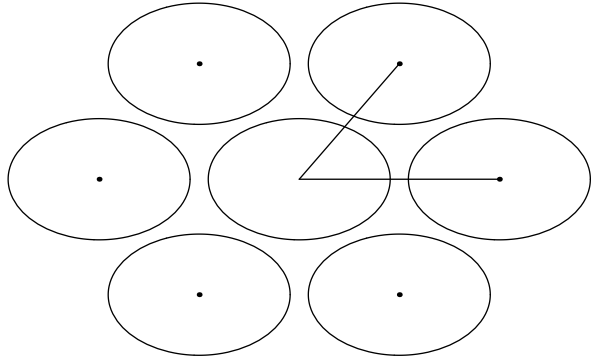}
\caption{\label{figellipsepack} Arrangement of densely packed ellipse-like aggregates.}
\end{figure}
\begin{figure}
\includegraphics[scale=1]{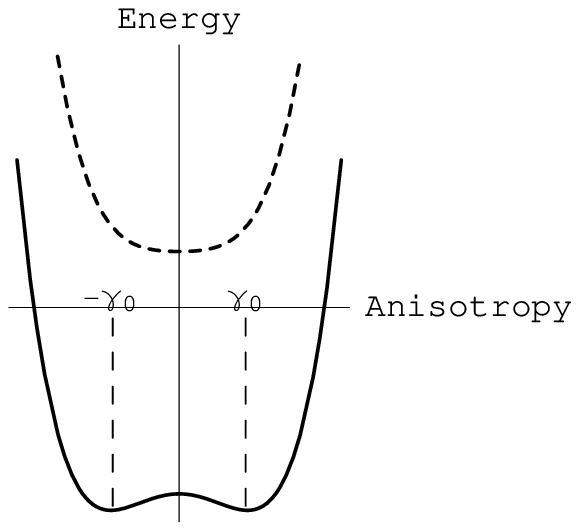}
\caption{\label{figeffenergy} Energy per aggregate (Eq.  (\ref{entrial})) as a function of the trial anisotropy parameter $\gamma$ for the charge density below the threshold (dashed) and above  the threshold (solid).}
\end{figure}

\begin{thebibliography}{99}
\bibitem{geom}
A. L. Kholodenko and T. A. Vilgis, Phys. Rep. {\bf 298}, 254 (1998).
\bibitem{wolfgang} J. Wolfgagng, S.M. Risser, S. Priyardshy, and D.N. Beratan,
J. Phys. Chem {\bf 101}, 2986 (1997).
\bibitem{feitel} J. Feitelson and G. McLendon, Biochemistry,
{\bf 30}, 5051 (1991).
\bibitem{viduna} D. Viduna, K. Hinsen and G. Kneller, Phys. Rev. E {\bf 62},
3986 (2000).

\bibitem{peyrard} J.~J.-L.~Ting and M.~Peyrard, Phys. Rev.~E
{\bf 53}, 1011 (1996);
K. Forinash, T. Cretegny, and M. Peyrard,
Phys. Rev. E {\bf 55}, 4740 (1997).

\bibitem{polyakova}
T. B. Polyakova and R. A. Suris,
Solid State Commun. {\bf 77}, 825 (1991).
\bibitem{conw} E. M. Conwell and S.V. Rakhmanova, Proc. Natl. Acad. Sci. U.S.A. {\bf 97}, 4557 (2000).
\bibitem{yoo} K.H. Yoo {\it et.al} Phys. Rev. Lett. {\bf 87}, 198102 (2001).
\bibitem{alex}S.S. Alexandre {\it et.al} Phys. Rev. Lett. {\bf 91}, 108105 (2003).
\bibitem{curv1}
Yu. B. Gaididei, S. F. Mingaleev, and P. L. Christiansen,
Phys. Rev. E {\bf 62}, R53 (2000);
P. L. Christiansen, Yu. B. Gaididei, and S. F. Mingaleev,
 J. Phys.: Condens. Matter, {\bf 13}, 1181 (2001).
\bibitem {arch}J. F. R. Archilla, P. L. Christiansen,
S. F. Mingaleev, and Yu.  B. Gaididei,
J. Phys. A.: Math. Gen. {\bf 34}, 6363 (2001);
J. Cuevas, J.F.R. Archilla, Yu.B. Gaididei and F.R. Romero,
 Physica D {\bf 163}, 106 (2002);
 J. F. R. Archilla, P. L. Christiansen, and Yu. B. Gaididei,
Phys. Rev. E {\bf 65}, 016609 (2002).
 \bibitem{george}
R. Reigada, J.M. Sancho,  M. Ibanes, and G.P. Tsironis,
 J. Phys. A, {\bf 34}, 8465 (2001);
  M. Ibanes, J.M. Sancho, and G.P. Tsironis,
 Phys. Rev. E {\bf 65}, 041902 (2002).
 \bibitem{bopp} M. Bopp, A. Sytnik, T. D. Howard, R. J. Cogdell and R. M. Hochstrasser, Proc. Natl. Acad. Sci. USA, {\bf 96}, 11271 (1999).
 \bibitem{oijen} A. M. van Oijen, M. Ketelaars, J. K{\"o}hler, T. J. Aartsma, J. Schmidt, Science,{\bf 285}, 400 (1999).
 \bibitem{kete} M. Ketelaars, A. M. van Oijen, M. Matsushita , J. K{\"o}hler, J. Schmidt, and T. J. Aartsma, Biophysical J. {\bf
 80}, 1591 (2001).
 \bibitem{matsu} M. Matsushita, M. Ketelaars, A. M. van Oijen, , J. K{\"o}hler, T. J. Aartsma, and J. Schmidt,  Biophysical J. {\bf
 80}, 1604 (2001).
 \bibitem{hong} X. Hong, Yu-X. Weng, and M. Li,   Biophysical J. {\bf
 86}, 1082 (2004).
 \bibitem{saxena-1998}
A. Saxena, R. Dandoloff, T. Lookman,
Physica A {\bf 261}, 13 (1998).

\bibitem{saxena-2000}
Y. Jiang, T. Lookman, and A. Saxena,
Phys. Rev. E {\bf 61}, R57 (2000).
\bibitem{ming02}
S.F. Mingaleev, Yu. B. Gaididei,
P.L. Christiansen, Yu. S. Kivshar,
Europhys. Letters, {\bf 59}, 403 (2002).
\bibitem{chaubal} C. V. Chaubal, L. G. Leal, J. Polym. Science B, {\bf 37},281
(1999).
\bibitem{abr} M.~Abramowitz and I.~Stegun, {\em Handbook of 
Mathematical Functions} (Dover Publications, Inc., New York, 1972).

\bibitem{dermott} G. McDermott {\it et al.} {\it Nature}, {\bf 374} 517 (1995).
\bibitem{bahatyrova} S. Bahatyrova {\it et al.} J. Biol. Chemistry, {\bf 279}, 21327 (2004).
\bibitem{gayberne} J.G. Gay and B.J. Berne, J. Chem. Phys. {\bf 74}, 3316 (1981).
\end{thebibliography}
\end{document}